\documentclass[useAMS,usenatbib]{mn2e}
\usepackage{graphicx,fleqn}
\DeclareGraphicsExtensions{.eps,.ps,.eps.gz,.ps.gz,.eps.Z}
\def \apj{ApJ}
\def \apjl{ApJL}

\def \aap{A\&A}

\def \mnras{MNRAS}

\input psfig.sty

\title[{\it Swift} and ground-based observations of GRB~050401.]{{\it
Swift} and optical observations of GRB~050401}

\author[]{Massimiliano De Pasquale$^{1}$, Andy P. Beardmore$^{2}$,
S.D. Barthelmy$^{3}$, P. Boyd$^{3}$, \newauthor D.N. Burrows$^{4}$,
R. Fink$^{3}$, N. Gehrels$^{3}$, S. Kobayashi $^{4,5}$, K.O. Mason
$^{1}$, \newauthor R. McNought$^{6}$, J. A. Nousek$^{4}$, K.L. Page
$^{2}$, D.M. Palmer$^{7}$, B.A. Peterson$^{6}$, \newauthor P.A.
Price $^{8}$, J. Rich$^{6}$, P. Roming $^{4}$, S.R. Rosen $^{1}$, T.
Sakamoto $^{3}$, B.P. Schimdt$^{6}$, \newauthor J. Tueller$^{3}$
A.A. Wells$^{2}$, S. Zane$^{1}$, B. Zhang $^{9}$, H. Ziaeepour$^{1}$. \\
$^{1}$Mullard Space Science Laboratory, University College London,
Holmbury St. Mary, Dorking Surrey, RH5 6NT, UK; mdp@mssl.ucl.ac.uk \\
$^{2}$University of Leicester, University Rd, Leicester, LE1
7RH, UK\\
$^{3}$ NASA/Goddard Space Science Flight Center, Greenbelt, MD
20771, USA\\
$^{4}$Department of Astronomy and Astrophysics, Pennsylvania State
University, 525 Davey Laboratory, University Park, PA 16802 USA\\
$^{5}$ Center for Gravitational Wave Physics, Pennsylvania State
University, University Park, PA 16802, USA\\
$^{6}$ Research School of Astronomy and Astrophysics, Mount Stromlo
Observatory, Cotter Road, Weston Creek, ACT 2611 Australia\\
$^{7}$ Los Alamos National Loboratories, Los Alamos, NM87545 USA\\
$^{8}$ Institute for Astronomy, 2680 Woodlawn Drive, Honolulu,
HA96822, USA\\
$^{9}$ Department of Physics, University of Nevada, Las Vegas, NV89154, USA }

\begin{document}

\date{Accepted...Received...}

\maketitle

\label{firstpage}

\begin{abstract}
We present the results of the analysis of $\gamma$-ray and X-ray
data of GRB~050401 taken with the {\it Swift} satellite, together
with a series of ground-based follow-up optical observations. The
{\it Swift} X-ray light curve shows a clear break at about 4900~s
after the GRB. The decay indices before and after the break are
consistent with a scenario of continuous injection of radiation from
the `central engine' of the GRB to the fireball. Alternatively, this
behaviour could result  if ejecta are released with a range of
Lorentz factors with the slower shells catching up the faster at the
afterglow shock position. The two scenarios are observationally
indistinguishable.

 The GRB~050401 afterglow is quite bright in the X-ray band, but weak
in the optical, with an optical to X-ray flux ratio similar to those
of `dark bursts'. We detect a significant amount of absorption in
the X-ray spectrum, with N$_{H}=(1.7 \pm 0.2) \times
10^{22}$~cm$^{-2}$ at a redshift of $z=2.9$, which is typical of a
dense circumburst medium. Such high column density implies an
unrealistic optical extinction of 30 magnitudes if we adopt a
Galactic extinction law, which would not be consistent with the
optical detection of the afterglow. This suggests that the
extinction law is different from the Galactic one.

\end{abstract}


\section{INTRODUCTION}
\label{intro}

Observations of Gamma-Ray Bursts (GRBs) have shown that they are
followed by fading X-ray, optical and radio afterglows. These are
thought to arise when the burst ejecta interact with the surrounding
medium and produce a shock, which propagates in the medium and heats
the electrons. The latter, cooling by synchrotron emission, produce
the observed radiation. Studies of afterglows can provide invaluable
information on the central engine of GRBs and on the circumburst
medium, and can possibly distinguish different subclasses in the
GRBs population.

 Some authors (see e.g. \citealt{laz02} and
references therein; \citealt{ber05} and \citealt{lamb05} for recent
results) have pointed out the existence of a subclass of GRB,
whose optical emission is at least $\sim2$ magnitude
below that of the average of the optically detected bursts. Different
models have been proposed to explain these ``dark bursts", ranging
from a cosmological origin (\citealt{bl02,fru99}) to scenarios where
they go off in relatively dense and highly absorbed regions (Lazzati
et al. 2002). It is also possible that many dark bursts may  be intrinsically
weak sources, the faint tail of the GRBs luminosity distribution
(\citealt{dp03}), or sources with a very rapid decay
(\citealt{gr98}) in the optical band.\\

The afterglow emission is seen to decay over time. In some bursts, a
clear light curve steepening is observed after an interval of order
days. The break is achromatic and typically attributed to the fact
that the energy release is collimated in a jet. Other irregular
temporal features are sometimes seen in bursts (see \citealt{zm02},
Zhang et al. 2005, Burrows 2005 and references therein): examples
include a rebrightening in GRB~970508, wiggles in GRB~020104 and
step-like features in GRB~030329. Also, ``bump" features have been
observed in several cases (e.g. GRB~970228, GRB~970508, GRB~980326,
GRB~000203C) and various interpretations have been proposed, e.g.,
``refreshed shocks" (\citealt{pan98}), supernova components
(\citealt{bloo99}, \citealt{rei99}, \citealt{gal00}), dust echoes
(\citealt{es00}) or microlensing (\citealt{gar00}). On the other
hand, signatures in the GRB lightcurve detected at earlier times may
provide diagnostic information about the nature of the injection and
eventually probe whether the energy is released impulsively during
the event or more continuously during the immediate post-burst epoch
(see e.g. \citealt{zm02}).

 Until recently, most follow-up observations did not start until a
few hours after the GRB, when the afterglow had already faded
significantly. This situation has changed with the launch of the
{\it Swift} mission, which provides both a rapid alert of GRB
triggers to ground-based observers, and rapid X-ray and optical/UV
follow-up observations of the burst afterglow. The  {\it Swift}
observatory (\citealt{ge05}) carries three science instruments: the
Burst Alert Telescope (BAT; Barthelmy et al. 2005), which locates
GRBs with 3' accuracy, the X-ray telescope (XRT, \citealt{ba05}) and
the Ultra-Violet Optical Telescope (UVOT, \citealt{rom05a}). When
BAT detects a GRB trigger, {\it Swift} slews towards the source
position within a few tens of seconds. Therefore, observations with
the {\it Swift} instruments yield high quality data and cover the
poorly investigated epoch occurring minutes after the burst.
Interestingly, many GRBs localized and observed by {\it Swift} have
shown no optical counterpart, even when optical observation started
100-200~s after the GRB onset. This provides evidence for a
population of ``intrinsically" dark GRBs (\citealt{rom05b}).

 In this paper, we report the properties of the optically faint {\it
Swift} GRB~050401, and discuss them in the light of the current models and
scenarios of GRBs.

\section{Analysis of the $\gamma$-ray and X-ray data.}

\begin{figure*}
 \includegraphics[angle=-90,scale=0.7]{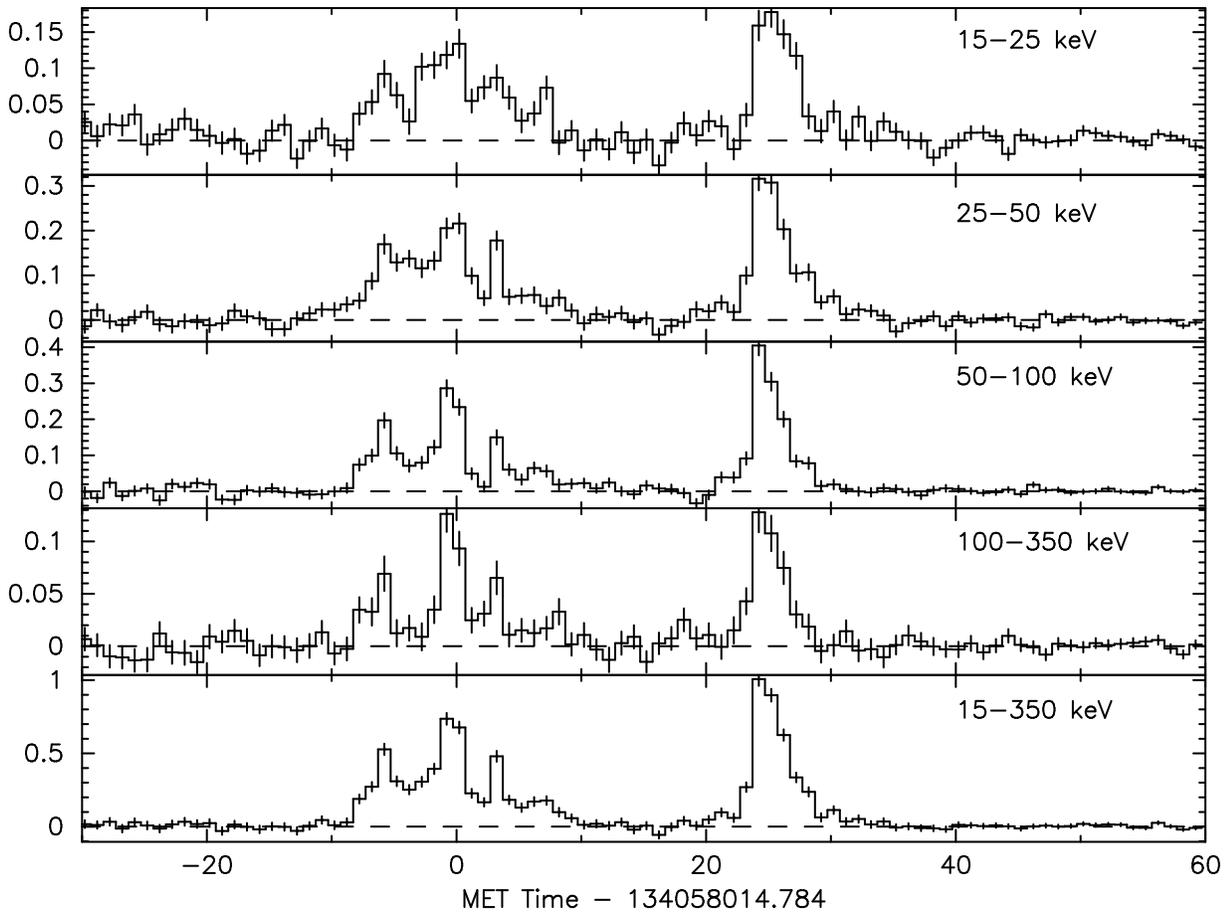}
 \caption{GRB~050401 light curve in Gamma-Rays.
On the Y-axis we show the background subtracted count/s per fully
illuminated detector for an equivalent on-axis source.}
 \label{f1}
\end{figure*}

 GRB~050401 triggered the BAT instrument at 14:20:15 UT on April 1,
2005 (\citealt{bar05}, \citealt{ang05}). The refined BAT position is
RA=$16^{h} 31^{m} 16^{s}$, Dec=$2^{\circ} 11' 35''$ with a position
uncertainty of 3' (95\% C.L., \citealt{sak05}). The
$\gamma$-ray band lightcurve started 9~s before the BAT
trigger time and it shows 4 main peaks (see Figure~\ref{f1}). The
peak count rate was 5000 counts/s (\citealt{bar05}).

Analysis of the BAT data (15-350~keV energy band) yields a GRB duration t$_{90}=33$~s.

We use the mask-weighted technique to subtract the background in the
BAT for spectral analysis, which is only effective up to 150 keV.
{\it Swift} began to slew towards the source about 25 sec after the
trigger, while the prompt emission was still active. We created
separated BAT spectra and response matrices for the pre-slew and
slew phases, and we fitted them jointly (see figure 2) with a simple
powerlaw model. No significant improvement in $\chi^{2}_{\nu}$ is
found with a cutoff power-law model. Results are reported in the
first entry of Table~1.\footnote{Throughout this paper, we report
errors at 1$\sigma$, unless otherwise indicated}.

The fluence detected by BAT in the $15-150$~keV range is
$(8.6\pm0.3) \times 10^{-6}$ erg cm$^{-2}$. Assuming a redshift of
$z=2.9$ (see later) and a spherically symmetric emission, this
corresponds to a $\gamma$-energy release of $9.6\times 10^{52}$ erg
between 15 and 150 keV in the cosmological rest frame of the burst
(derived by means of the 'k-correction' of Bloom et al.~2001). This
result differs from that obtained by Chincarini et al.~(2005), who
reported an energy spectral index $\beta=-0.13\pm0.05$ (using the
20-150 keV data) and a  $\gamma$-energy release ($\sim2.8\times
10^{53}$ erg) in the 15-350 keV band. However, we note that
GRB~050401 prompt emission was also detected by Konus-Wind
(\citealt{go05}). The Konus data (see below) suggest a steepening of
the spectral index above E$_{0}\sim$ 150 keV, which might explain
the large difference between the two analyses.

 As observed by the Konus-Wind instrument, GRB~050401 had a
duration of $\sim36$~s and a fluence of $(1.93\pm0.04) \times
10^{-5}$ erg cm$^{-2}$ in the 0.02-20 MeV band. \cite{go05} analyzed
the spectrum gathered in the first 3 peaks (0-17~s after the
trigger, first segment) and last peak (24.8-32~s after the trigger,
second segment) separately (see again Table~1). We note that, if we
adopt a cutoff powerlaw-model and fix the break energy $E_0=150$~keV
(as inferred from the Konus data), the spectral index obtained by
\textit{Swift}, which is averaged over the whole gamma-ray emission
phase, is consistent with that obtained by Konus.

Observations were not possible with the {\it Swift} UVOT telescope
because of the presence of a 4$^{th}$ magnitude star in the field of
view, while XRT started observations about 130~s after the trigger. An
unknown bright X-ray source was detected at R.A. = $16^h$
$31^m$ $29^s$, Dec = 02$^{\circ}$ 11' 14'', within 42 arcseconds of
the initial BAT position (\citealt{ang05}); the coordinates were
later confirmed by ground processing. This source subsequently
faded, indicating that it was the X-ray counterpart of GRB~050401.

 The first data were taken with the XRT (\citealt{hi05}) in its
Imaging Mode (IM), then followed by a segment of data in the
PhotoDiode mode (PD). After that, because of the bright star near
the edge of the XRT field of view, the detector continuously
switched between Windowed Timing (WT) and Photon Counting (PC) mode.
The initial PC mode data were piled-up and we did not include them
in the analysis. In the first day of observation, {\it Swift}
observed the X-ray afterglow until 5.6 hours after the trigger. More
follow-up observations were performed 4.43-6.38 and 6.48-12.33 days
after the trigger. The total exposure time is about 48~ks, divided
into 21~ks in WT mode and 27~ks of PC mode (including piled-up
data).

 Analysis of XRT data was performed using the XRT pipeline software.
The accumulated DN in the IM data were converted to a count-rate
following the method of \citealt{hi05}. The PD mode data points have
been obtained by subtracting the contribution of the corner
calibration sources. As for the WT mode data, the extraction regions
for the source and for the background consist of boxes of 60x40
pixels. The source and background extraction regions for the PC data
are circles of 20 pixel and 60 pixel radius, respectively. We
considered the 0-2 grade events for the WT mode and 0-12 for the PC
mode, which cover events up to 4 pixels in size. For the spectral
analysis, we generated the physical ancillary response matrices with
the task \textit{xrtmkarf}, while the response matrices (RMs) were
retrieved from the latest {\it Swift} calibration database, CALDB
20050601 (http://heasarc.gsfc.nasa.gov/docs/heasarc/caldb/swift/).
At the time of writing, these RMs are considered as the most
reliable, basing on comparison with spectra of  calibration sources.
We only considered  data taken with CCD detector temperature T$<-48$
C.

 We restricted both the spectral and temporal analysis to counts
within the 0.4-10 keV band since, at the time of writing, a good
calibration of both WT and PC data is available in this band.
Data were rebinned in energy by requiring a minimum of 15 counts per bin.

The X-ray lightcurve is shown in Figure~\ref{f3}. The presence of a
break is clearly evident: if we try to fit the whole lightcurve with
a single powerlaw, we get an unsatisfactory result
($\chi^{2}_{\nu}=904/114$), while the use of a broken powerlaw model
($A t ^{\alpha_1}$ for $t\leq t_{b}$; $A t_{b}^{\alpha_{1} -
\alpha_{2}} t^{\alpha2}$ for $t\geq t_{b}$) gives a statistically
significant improvement ($\chi^{2}_{\nu}=129/112)$. In this case,
the best fit parameters are: decay indices $\alpha_{1}=-0.63\pm0.02$
and $\alpha_{2}=-1.46\pm{0.07}$ (before and after the break,
respectively) and break time $t_{b}=4900\pm 490$~s\footnote{We note
that at the time of the slope change \textit{Swift} was not
observing the GRB.}. The break time is consistent with that reported
by Chincarini et al.~(2005), who report $\sim 1300$~s corrected for
cosmological time dilation, i.e. their break time has been divided
by $(1+z)$. We also tried to fit the lightcurve with a smoothly
joined broken power law model (\citealt{beu99}), to determine the
''sharpness'' of the X-ray lightcurve break. However, we found that
data do not allow us to discriminate between a smooth and a sharp
transition.

A fit of the WT spectral data taken before the lightcurve break time
with an absorbed power-law model reveals a considerable amount of
absorption, N$_{H}^{tot}=(1.5\pm0.1) \times10^{21}$~cm$^{-2}$,
clearly in excess of the Galactic value reported by Dickey \&
Lockman~(1990), $N^{Gal}_H=4.85\times10^{20}$~cm$^{-2}$, and the
value inferred from the Galactic extinction map of Schlegel et
al.~(1998), N$_{H}^{Gal}=3.6\times10^{20}$~cm$^{-2}$ (in this case
the extinction, A$_{V}=0.2$ has been converted to N$_{H}^{Gal}$
using results by Predehl and \& Schimtt 1995).

In order to better estimate the column density corresponding to the
circumburst medium only, we  repeated the fit by accounting
separately for the Galactic and extragalactic absorption columns,
the latter rescaled at $z=2.9$. We fixed the Galactic absorption to
the Dickey \& Lockman (1990) value. This gives an extragalactic
column density of N$_{H} \equiv N_H^{tot} - N_H^{Gal} \approx (1.7
\pm0.2) \times 10^{22}$~cm$^{-2}$ (see Table~1; spectrum and best
fit model are shown in Figure~\ref{f4}). To assess the robustness of
this detection, we then repeated the fit without adding the
extragalactic component, obtaining a $\chi^{2}_{\nu}=405/262$. This
means that accounting for the extra absorption N$_{H}$ produces a
statistically significant improvement. The F-test has also been
widely used to test the significance of a spectral component,
although, when applied to parameters such as N$_{H}$ (which is
bounded to be $>0$) this may be inappropriate in a strict sense (see
\citealt{prot02}). For completeness, we report that we checked the
F-test statistic, obtaining a value of 117 with a probability that
the improvement is due to chance of $\sim 10^{-22}$.

We find no evidence for spectral evolution: parameters consistent
with those given above are obtained when fitting the spectra taken
before and after the $~4900$~s break. An analysis of the PC spectrum
with the same spectral model also does not show any significant
difference in the spectral parameters (see Table~1). When compared to
the X-ray afterglows detected by other observatories, like {\it
BeppoSAX}, {\it Chandra \/} and {\it XMM-Newton}, the afterglow of
GRB~050401 appears to be a moderately bright source: the 1.6-10 keV
X-ray flux normalized at 11 hours after the burst is $(2.3\pm0.2 )
\times 10^{-12}$ erg cm$^{-2}$s$^{-1}$ (see \citealt{ber03},
\citealt{rom05b}, De Pasquale et al. 2005).


\section{Follow-up observations in different energy bands}

 Although the {\it Swift} UVOT could not observe at the GRB position, a
series of ground-based optical follow up observations was performed,
triggered by the prompt {\it Swift} localization.

 The afterglow was first identified by \cite{ryo05}, who detected it
in images taken shortly after the GRB. Later, following the
distribution of the BAT trigger, the BAT error circle was imaged with the
40-inch telescope at Siding
Spring Observatory. These observations consisted of two unfiltered
120~s exposures, followed by 41 $R$-band 240~s exposures. Data
showed a faint fading source, at coordinates RA=16:31:28.81,
Dec=+02:11:14.2 (J2000), i.e. within the XRT error circle. This
object was not present on the Digitised Sky Survey plates
(\citealt{mcn05}).

 The rapid distribution of the afterglow position, along with a
finding chart, enabled other follow-up observations including some
spectroscopic ones. \cite{fyn05} reported the detection of a system
of absorption lines in the optical spectrum detected with FORS2 at
the Very Large Telescope, indicating a redshift of $z=2.9$. Other
observations followed, up to $\sim0.5$ days after the GRB onset. The
full list is summarized in Table~2 and the optical light curve is
shown in {Figure~\ref{f3}}. We found that the optical decay law is
not consistent with a single powerlaw ($\chi^{2}=24.8$ for 13
d.o.f.). We caution that the optical light curve has been obtained
by observations performed by different facilities, which used
diverse calibrations. In particular, the photometry of all exposures
taken with the Siding Spring Observatory is relative to the first
$R$-band image, which was then calibrated to match the USNO-A2.0
catalogue red magnitude. Because of this, although the relative
magnitudes are accurate, the overall normalization can only be
considered to be accurate at the $\sim 0.5$~mag level. ARIES
measurements also refer to USNO-A2.0 stars, while the Maidanak
estimate is based on stars of the similar USNO-B2.0 catalogue, and
D'Avanzo et al. 2005, and Greco et al.~2005 data used the Landolt
catalogue. Therefore some systematic bias may well be present.
However, taken at face value the combined data suggest a
non-monotonic decaying behaviour. We note that the decay slope in
the first 10000~s (first part of the Siding Spring data), is
$\alpha_{0}=-0.68\pm0.06$, consistent with the first X-ray decay.
The Siding Spring data  are not consistent with being constant. A
powerlaw fit to these data in isolation gives $\alpha=0.56\pm0.2$.
The case $\alpha=0$ can be rejected at the 2.5$\sigma$ confidence
level.

\section{Discussion.}

\subsection{The break in the X-ray light curve.}

 As we have seen, the X-ray light curve of GRB~050401 registered with XRT shows
an initial decay slope of $\alpha_{1}=-0.63\pm0.02$, which steepens
at $\sim 4900$~s. After this time, the source fades with a slope of
$-1.46\pm0.07$.

In principle, there are several physical mechanisms that can dictate
the physical behavior of the afterglow during its initial decline
and can produce a break at these early times. For instance, as
described by \cite{sa99} and \cite{rh99}, a break in the light curve
is expected if the fireball outflow is not spherically symmetric but
collimated within a jet. Such an effect has been invoked to explain
breaks in Gamma-Ray Burst afterglow light curves in several cases
(see, for example, \citealt{sa99}, \citealt{pk02}, \citealt{fr03}).
However, we suggest that this explanation is not applicable to the
case at hand, for at least two reasons. First, according to the
fireball model of GRB afterglows, the observed temporal decay index
and spectral slope should be linked through the so-called closure
relation (see e.g. \citealt{pri02}). This relation depends on the
kind of expansion (spherical or jet), on the density profile of the
medium (uniform or wind) and on the cooling state of the electrons
responsible for the synchrotron emission. In particular, for
jet-like emission it should be $\alpha=2\beta-1$ or $\alpha=2\beta$
depending on whether the cooling frequency $\nu_{c}$ is above or
below the X-ray frequency  $\nu_{X}$ (see \citealt{sa99}). In the
case at hand, we find $\beta=-0.9\pm0.03$ (the spectral index is not
supposed to change after the beginning of the jet expansion phase)
and $\alpha=-1.42\pm0.07$ (decay index after the break), so none of
the above closure relations is satisfied. Second, if interpreted
within the jet break scenario the observed parameters are not
compatible with the relation between the peak energy and the
(collimation corrected) $\gamma$-ray energy release proposed by
\cite{ghi04}. In fact, by using $E_{\gamma, iso}=3.5\times10^{53}$
erg (as inferred from Konus data) and $t_b=4900$s we obtain a jet
opening angle $\theta\simeq1^{\circ}$ (see expression~1 in
\citealt{ghi04}). In turn, this corresponds to a collimation
corrected energy $E_{\gamma}=4.8\times10^{49}$~erg and to a peak
energy in the cosmological rest frame of $E_{peak} =56$ keV, in
clear disagreement with the value inferred by Konus data ($\sim 500$
keV).

Another potential reason for a break in the light curve is that the
energy release is spherically symmetric, but the X-ray observation
occurs while $\nu_{C}$ passes through the X-ray band, causing the
light curve to steepen by 0.25 (see \citealt{sa99}). However, in
this case we should also observe a steepening of $0.5$ in the
spectral slope while there is no evidence for that across the break
of GRB~050401. Moreover, a change of 0.25 is not sufficient to
account for the steepening in the light curve.

On the other hand, we find that both the initial shallow decay and
the break can be explained by a model in which the central engine
continues to inject radiative energy into the fireball for several
thousands of seconds. This scenario has been investigated by
\cite{zm01} and \cite{zm02}. By assuming a source luminosity law of
the kind $L \propto t^{q}$, where $t$ is the intrinsic time of the
central engine (or the observer's time after the cosmological time
dilation correction), these authors found that continuous injection
of energy influences the fireball and the observed light curves as
long as $q>-1$. In this case, the spectral and decay slopes are
linked through the relation:
\begin{equation}\label{1}
    \alpha=(1-q/2) \beta + q + 1 \, ,
\end{equation}
\noindent which holds if the observed X-ray frequency is between
the synchrotron peak frequency and the cooling frequency (see
later). By using the observed values of $\alpha$ and $\betaÂ$, we
obtain $q\sim-0.5$ and $q\sim-1$ before and after the break time,
respectively. The latter is
consistent with no
injection (since $q<-1$ does not influence the fireball dynamics),
while before the time break the central engine injects energy with a
luminosity law $L \propto t^{-0.5}$. The change in the decay
slope occurs at the point when the central engine ceases to inject significant
amounts of energy. We note that the decay slope of the optical and
the X-ray flux before the breaks are consistent within the errors. This is
in agreement with the continuous injection model, as long as the
optical and the X-ray band belong to the same spectral segment.

There is variant to this scenario, which is observationally
undistinguishable, i.e. a model in which the central engine
activity is as brief as the prompt emission itself but, at the end of the
prompt phase, the ejecta are released with different velocities
(Lorentz factors, see Panaitescu et al.~2005). The fastest shells
initiate the forward shock, decelerate, and are
successively caught by the slowest shells. The consequent
addition of energy in the blast-wave mitigates the deceleration and
the afterglow decay rate. Assuming that the mass $M$ of the ejecta
follows the law

\begin{equation}\label{2}
    M(>\gamma) \propto \gamma^{s} \, ,
\end{equation}
where $\gamma$ is the Lorentz factor, one can find an effective $s$
value that mimics the effect of non vanishing $q$ index in the luminosity
law. By following  Zhang et al.~(2005), this is:
\begin{equation}\label{3}
    s=-(10+7q)/(2-q) \, ,
\end{equation}
therefore, the value $q=-0.5$ inferred above is equivalent to an
$s$-index of $s=-2.6$. This explanation has been proposed, for
example, to explain the initial mild decline of the optical light
curve of GRB~010222, which shows a decay slope of $\alpha_{O}=-0.7$
(Stanek et al. 2001, Bjornsson et al. 2002) for $\sim10$~hours after
the trigger, followed by a break and a steeper decay. Bjornsson et
al. (2004) also proposed that injection of energy by slow shells
could explain the wiggles in the light curve of GRB~021004.

In the case of GRB~050401, it is noteworthy that a possible optical
rebrightening seems to take place shortly after the change of the
slope of the X-ray. In the framework of the continuous energy
injection model, this could be explained by the onset of a "reverse
shock". The basic idea is that, after the end of the injection
phase, a reverse shock wave crosses the whole ejecta, heating them
and causing a peak in the emission. After that, the shocked ejecta
start to cool adiabatically once again. However, in order to assess
this issue a more detailed model investigation is required, which is
beyond the scope of this work.

As previously discussed, the analysis of the X-ray data taken after
the break is consistent with a scenario in which a "standard"
fireball expands in a constant density medium, provided that the
observed X-ray band lies between the synchrotron peak frequency and
the cooling frequency. Following \cite{sa99}, the
closure relation should in this case read $\alpha = (3/2) \beta $, which is
satisfied within $1\sigma$. One possible complication is that, if
the fireball expands in a medium with constant density, the cooling
frequency is expected to decrease with time according to $\nu_{C}
\propto t^{-1/2}$. Accordingly, in most X-ray afterglows, the
cooling frequency is already between the peak frequency and the
X-ray observing frequency less than 1-2 days after the GRB onset. In
contrast, our data seems to suggest that in the case of GRB~050401
the cooling frequency remains above the X-ray frequency for about
$10^{6}$~s. While this may be explained in terms of relatively low
values of magnetic field energy ($\epsilon_{B}$) and density (see
Sari et al. 1999), we also note that a transit of the cooling
frequency through the X-ray band after 20,000~s cannot be completely
excluded by our data: the change in the decay slope and in the
spectral slope would be 0.25 and 0.5 respectively, and hence
difficult to detect due to very low statistics in the late time XRT
detections of the afterglow.

\subsection{The optical and X-ray properties of GRB~050401.}

 The GRB~050401 afterglow is quite bright in the X-ray band, but weak in
the optical, with an optical to X-ray flux ratio similar to those of
'dark bursts'. \\
In order to compare its properties with those typically observed in
other GRBs, we show in figure~\ref{f5} the relation between the
optical and X-ray fluxes for a series of GRB afterglows detected by
{\it BeppoSAX}. As pointed out by several authors (e.g.
\citealt{dp03}, \citealt{rom05b}), the GRBs with optical
counterparts exhibit a correlation between the fluxes in these two
spectral bands. Several GRBs without optical counterparts also show
an X-ray emission consistent with that expected by assuming the
validity of the same correlation, indicating that they may well be
simply the faint tail of the same population. However, there is
evidence for a sample of dark GRBs which have "normal" X-ray fluxes
but with tight upper limits in the optical. Jakobbson et al. (2004)
reached similar conclusions by comparing the spectral index between
the optical and the X-ray band, $\beta_{OX}$, with the expectations
of the fireball model, which requires $\beta_{OX}\leq-0.5$. They
found that at least 10 \% of the events in their sample had
$\beta_{OX} > -0.5$, and called them the \textit{truly} dark GRBs.
To explain the optical faintness of these bursts, two main scenarios
have been proposed. The first idea is that they occur at very high
redshift, possibly following the death of Population II and III
stars (\citealt{bl02}), in which case the optical flux is washed out
by the intervening  Ly-$\alpha$ forest. The second idea is that dark
GRBs have lines of sight passing through large and giant molecular
clouds (hereafter GMCs). GMCs are rich in dust, which extinguishes
optical and UV light very efficiently.

 GRB~050401 appears to be an 'intermediate' case. An optical afterglow
is detected, but is very faint relative to its X-ray flux compared to other
GRBs with optical counterparts. Its optical to X-ray spectral index
is $\beta_{OX}=-0.33$, which makes this source a dark burst
according to Jakobsson et al. (2004) classification.
 Given the detection of an optical counterpart with a likely redshift
of $z=2.9$, we can exclude the hypothesis of a very high redshift.
On the contrary, the X-ray spectrum indicates a high absorption,
typical of GMCs (\citealt{rp02}, \citealt{gw01}). \\
Therefore, a natural question is: can the hypothesis of extinction
in this medium explain the weakness of the optical emission
detected? The light detected at Earth in the R band (centered at 700
nm) was emitted at a wavelength of 180 nm in the GRB cosmological
restframe at a redshift of $z=2.9$. The simplest working hypothesis
is to assume that the same medium is responsible for both the X-ray
and optical  extinction. In this case, the hydrogen column density
measured from the X-ray spectrum (N$_{H}=1.7 \times
10^{22}$~cm$^{-2}$) would correspond to an absorption in the V band
of $A_{V}=10$ magnitudes (\citealt{zom90}). Assuming a Galactic
extinction curve of $A_{\lambda} / A_{V} \propto \lambda^{-1}$, this
results in a predicted extinction of $A_{\lambda}\sim30$ magnitude
for $\lambda=180$~nm. This value is unreasonably high and would
imply that the optical afterglow was unrealistically
bright. Thus we can exclude this simple explanation.\\

Instead, comparing with the optical-to-X-ray flux ratio typical of
other 'non-dark' GRBs (\citealt{dp03}, \citealt{jak04},
\citealt{rom05b}) we could expect a plausible extinction
of about 3 magnitudes, which is clearly not in agreement with the
measured X-ray absorption when we adopt the
Galactic extinction curve. We note that a discrepancy like this has
been noted in several other cases (for a summary, see \citealt{str04}),
but for GRB~050401 we have the advantage of a fairly
constrained value of the absorption parameter.

In order to reconcile the value of absorption with the likely
extinction, a few hypothesis have been proposed. The first scenario
involves the presence of a gas-to-dust ratio much lower than the
Galactic one and/or a dust grain size distribution skewed toward
large grains (\citealt{str04}).
 This first case is, for instance, typical of dwarf galaxies like the
Small Magellanic Cloud (SMC). In fact, for the SMC interstellar
medium (ISM) a dust content $\sim1/10$ of the Galactic one has been
inferred (\citealt{pei92}). Thus, we would have $\sim 3$ magnitudes
of extinction based on the measured value of the X-ray absorption.
This value is close to that required. We note, however, that the
best-fit $N_{H}$ value has been obtained by assuming Galactic metal
abundances, while the metallicity of SMC is $1/8$ of the Milky Way
(\citealt{pei92}). If this low metallicity were adopted, we would
need to adjust the value of $N_{H}$ upwards by a factor of $\sim 7$,
given that the majority of the absorption in the X-ray band is
produced by heavy elements, so this is not a solution.
\\
Another scenario that could result in a low $A_{V}/N_{H}$ is that
there is a change in both the gas-to-dust ratio and the distribution
of grain size, with the latter enriched in large grains by the
effects of the high-energy radiation of the GRBs. Dust grains can be
heated and evaporated by the intense X-ray and UV radiation fields
up to $\sim20$ parsecs from the GRB (\citealt{wd00}, \citealt{dh02},
\citealt{fru01}). \cite{pe03} and  \cite{pl02} show that the consequence
of exposing dust to intense
radiation fields can be a grain size distribution flatter than the
original one. The main reason is that dust destruction is more
efficient on small grains. \cite{pe03} computed the
extinction curve that is obtained if standard Galactic dust is
exposed to a GRB and found that the extinction curve can be very
flat, at least for bursts lasting more than a few tens of seconds.

Finally, a distribution of grain size skewed toward large
grains can also be produced by an efficient mechanism of coagulation
of smaller grains in a dense environment (\citealt{kim96},
\citealt{ma01}), in which case the dust-to-gas ratio is unaffected.
Both of the above explanations could lessen the discrepancy between
the low UV extinction and the high X-ray absorption detected.

\section{Conclusions.}

We have presented {\it Swift} observation of the Gamma-Ray Burst
050401 and we have discussed the properties of the prompt emission,
and, in more detail, the X-ray afterglow. The light curve of this
burst shows a break 4900~s after the trigger, changing from a decay
index of $\alpha_{1}=-0.63$ to $\alpha_{2}=-1.46$, while the
spectral energy index does not change. To explain this behavior, we
have proposed that the 'central engine' has been active until the
time of the break, with luminosity described by the law $L\propto
t^{-0.5}$. Another possibility is that the central engine activity
turned off as the prompt emission ended, but the shells had a broad
distribution of Lorentz factors. In this case, the slowing front of
the GRB blastwave is continually re-energized by the arrival of
progressively slower shells and the flux decay is therefore
mitigated. We find that the decay slope observed before the break
may be reproduced if the shells were emitted with a powerlaw
distribution of Lorentz factor, $M \propto \gamma ^{-2.6}$.\\ The
peculiar behaviour of the optical light curve can be in agreement
with the two scenarios proposed.\\
After the break time, the profiles of the X-ray light curve and
spectrum are consistent with those expected when a fireball expands
in a circumburst medium with constant density, and the observed
X-ray band lies between the
synchrotron peak frequency and the cooling frequency.\\
 Even though the X-ray flux of the GRB~050401 afterglow is high, the
optical counterpart is faint. This leads to a low optical to X-ray
flux ratio similar to that of dark GRBs which are likely to be
obscured by some mechanism. The spectral analysis shows clear
evidence of absorption, namely N$_{H}=(1.7\pm0.2) \times
10^{22}$~cm$^{-2}$, at the redshift $z=2.9$ of the GRB. This value
is typical of giant molecular clouds where star forming regions are
located. The detection of a dense circumburst medium could lead us
to the conclusion that this "obscuration mechanism" is, at least in
this case, extinction.\\
 However, the amount of extinction extrapolated by assuming the
Galactic extinction law is far too high to be physically acceptable.
This may be evidence that the circumburst medium is characterized by
a dust grain size distribution different from the Galactic one, and
skewed towards large grains. This could be due either to coagulation
of smaller grains or to small dust grain destruction due to high
energy photons produced by the GRB. In the latter case the
dust-to-gas ratio would also be different from the Galactic one.

Acknowledgements: We are grateful to an anonymous referee for
his/her suggestions that led to a substantial improvement of the
draft. APB, KPA acknowledge support for this work at Leicester by
PPARC. SZ also thanks PPARC for support through an Advanced
Fellowships.


\clearpage

\begin{table*}
\begin{center}
\begin{tabular}{ccccccc}
\hline \\ Instrument &
Section & $\beta$ & $\beta_1$ & $E_0$ & N$_{H}$ & $\chi^{2}_{\nu}$ \\
        &   & &       & & $(10^{22})$ & \\
\hline \\
Swift: & & & & & \\
&  BAT Prompt Emission       & $-0.53\pm0.07$  &  &&             & 136/115
\\
&  WT pre-break   & $-0.9\pm0.03$   & && $1.7\pm0.2$   & 280/261 \\
&  WT post-break  & $-0.9\pm0.13$   & && $1.7$         & 16.3/21 \\
&  PC             & $-0.75\pm0.15$  & && $1.7$         & 21.8/18 \\
\hline
Konus-Wind: & & & & & \\
&  first segment & $-0.15\pm0.16$  & $-1.65\pm0.31$ & $156 \pm 45$
&   &  \\
&  second segment & $+0.17\pm0.21$  & $-1.37\pm0.14$ & $119 \pm 26$
&   &  \\
\hline
\end{tabular}
\caption{Values of the parameters for the spectral fit of
GRB~050401. Swift data have been fitted with a simple power-law
model (energy index $\beta$) and divided into four sections: prompt
emission, WT data before the break in the light curve, WT data after
the break, PC data. For the latest two sections, N$_H$ has been kept
fixed to the value obtained from the WT pre-break data. Errors are
at 68\% confidence level. For comparison, we also reported the
results of the fitting of Konus-Wind prompt emission data performed
by Golenetskii et al. (2005). A Band model has been adopted by these
authors, so in this case  $\beta$, $\beta_1$ are the low and high
energy indices and $E_0$ is the break energy.} \label{tab_spec}
\end{center}
\end{table*}

\vspace{5cm}

\begin{table*}
\begin{center}
\begin{tabular}{ccc}
\hline \\
Time after GRB
(days) & R magnitude & Reference \\
\hline \\
  0.040 & $21.05\pm 0.3$ & Siding Spring Observatory \\
  0.043 & $21.25\pm 0.3$ & ''\\
  0.051 & $20.85\pm 0.2$ & ''\\
  0.055 & $21.35\pm 0.3$ & ''\\
  0.063 & $21.85\pm 0.4$ & ''\\
  0.071 & $21.55\pm0.3$  & ''\\
  0.086 & $21.55\pm 0.2$ & ''\\
  0.102 & $21.85\pm0.3$  & ''\\
  0.126 & $21.85\pm0.3$  & ''\\
  0.149 & $21.85\pm 0.4$ & ''\\
  0.173 & $22.05\pm 0.3$ & ''\\
  0.24 & $21.27\pm0.2$ & Misra et al. 2005 (GCN 3175) \\
  0.39 & $21.97\pm0.2$ & Kahharov et al. 2005 (GCN 3174) \\
  0.46 & $22.47\pm0.4$ & Greco et al. 2005 (GCN 3319) \\
  0.47 & $22.95\pm0.1$ & D'Avanzo et al. 2005 (GCN 3171)\\
\hline
\end{tabular}
\caption{ Log of GRB~050401 Optical observations. Values quoted in
this table have been corrected for Galactic extinction. As explained
in the text, the zero-points for these observation might be
uncertain for $\sim0.5$ magnitude due to USNO-A2.0 calibration.}
\label{tab_obs}
\end{center}
\end{table*}

\clearpage

\clearpage

\begin{figure*}
\includegraphics[angle=-90,scale=0.7]{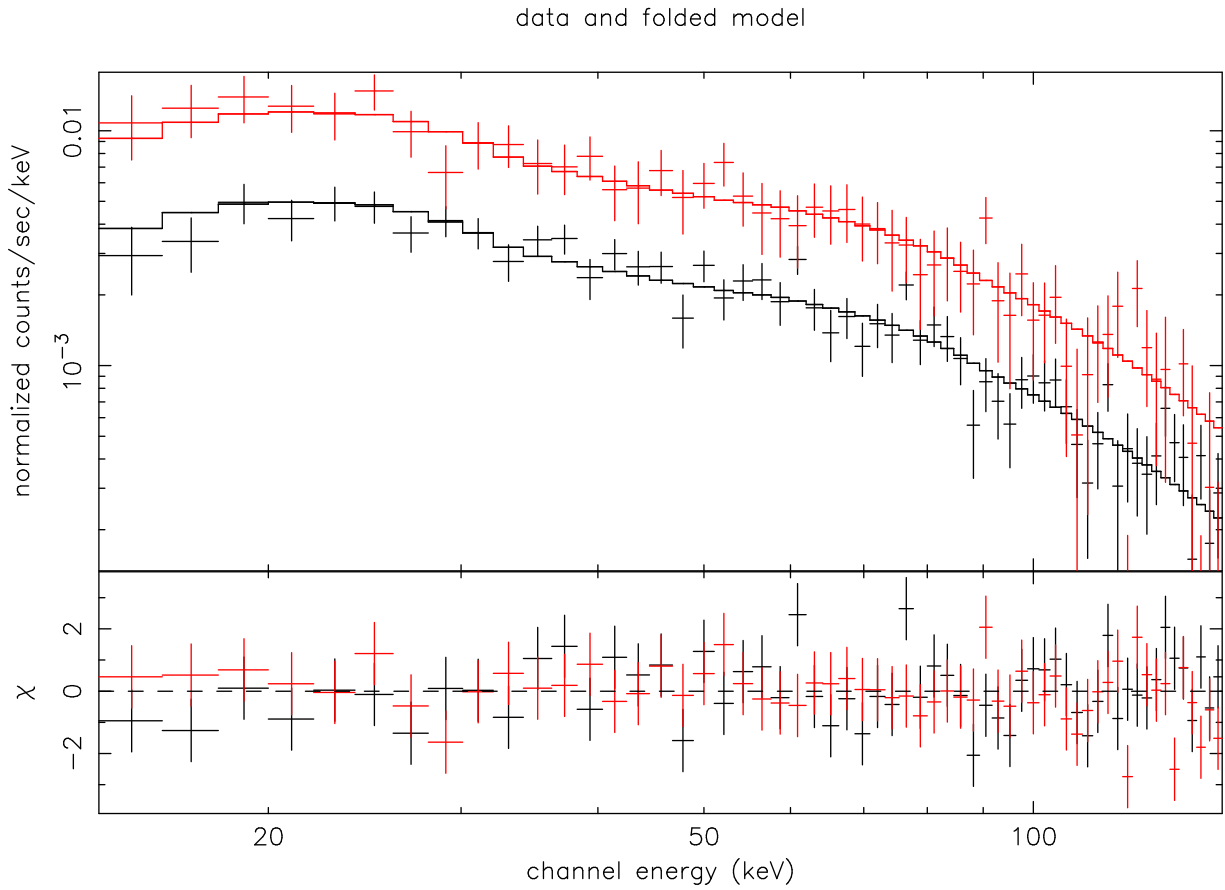}
\caption{GRB~050401 $\gamma$-ray spectrum detected by BAT before
(red colour) and during the slew (black). Both spectra are
consistent with the same fitting power law model, plotted as a solid
line (see first entry in Table~1).} \label{f2}
\end{figure*}

\clearpage

 \begin{figure*}
 \includegraphics[angle=-90,scale=0.6]{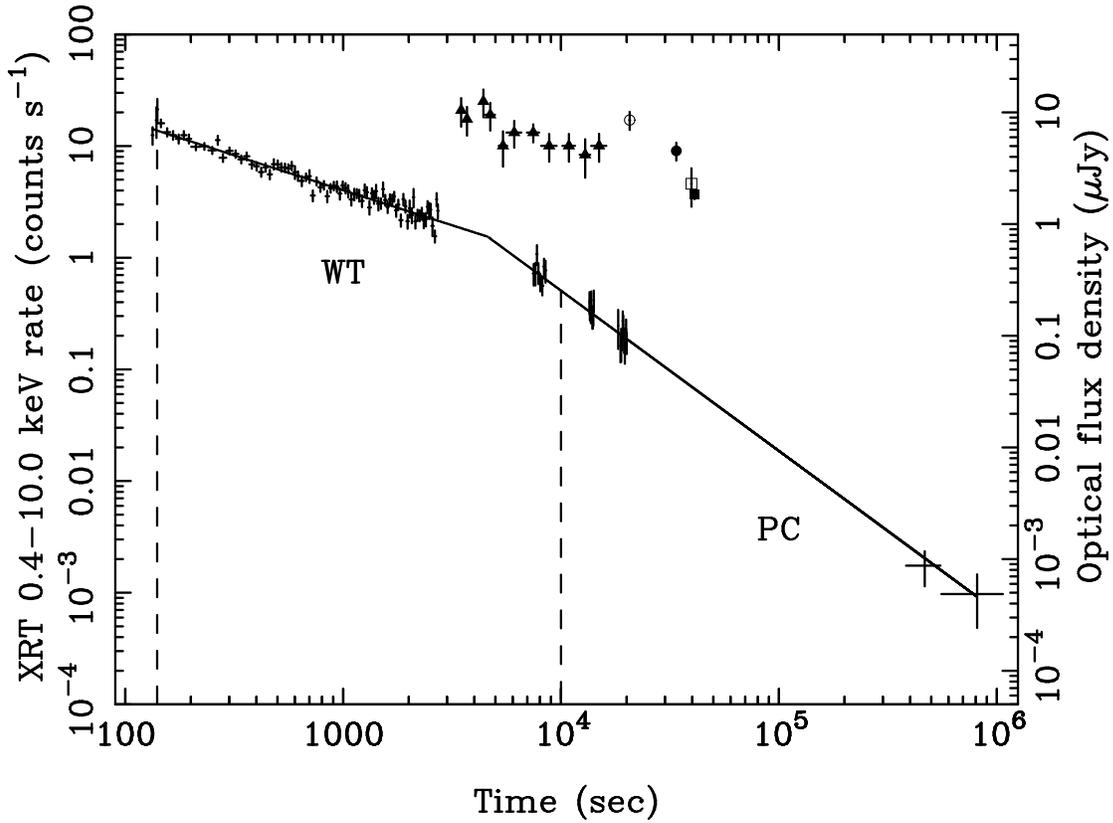}
\caption{GRB~050401 afterglow light curves. The crosses represent
the X-ray light curve registered with XRT (0.4-10~keV band). The
first point is taken in Image mode, followed by 2 Photodiode points.
After that, data have been taken in Windowed Timing (WT) mode from
0.14 up to 9~ks after the trigger (section between the two vertical
dashed lines), and in Photon Counting (PC) mode from 13.6 up to
1050~ks after the trigger. The optical afterglow light curve is
plotted in the upper part of the figure. Triangles: Siding Spring
data. Circle: ARIES (Misra et al.~2005) data. Filled circle: MAO
(Kahharov et al.~2005). Square: Bologna (Greco et al.~2005) data.
Filled square: TNG (D'Avanzo et al.~2005) data.} \label{f3}
\end{figure*}

 \begin{figure*}
 \includegraphics[angle=-90,scale=0.6]{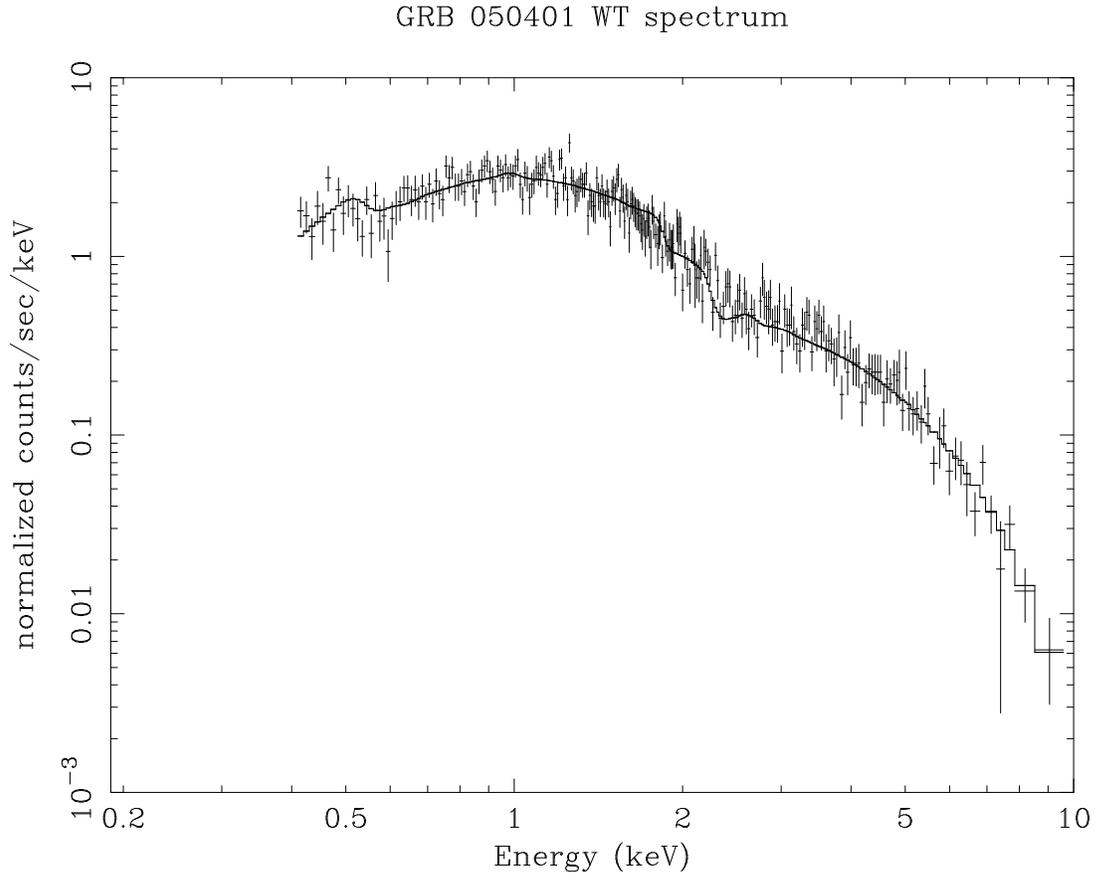}
 \caption{GRB~050401 X-ray spectrum registered with XRT in WT mode
 before the break. The solid line is the best fitting absorbed
power-law model (see text and Table~1 for details).}
\label{f4}
 \end{figure*}

\clearpage

\begin{figure*}
 \includegraphics[angle=0,scale=0.7]{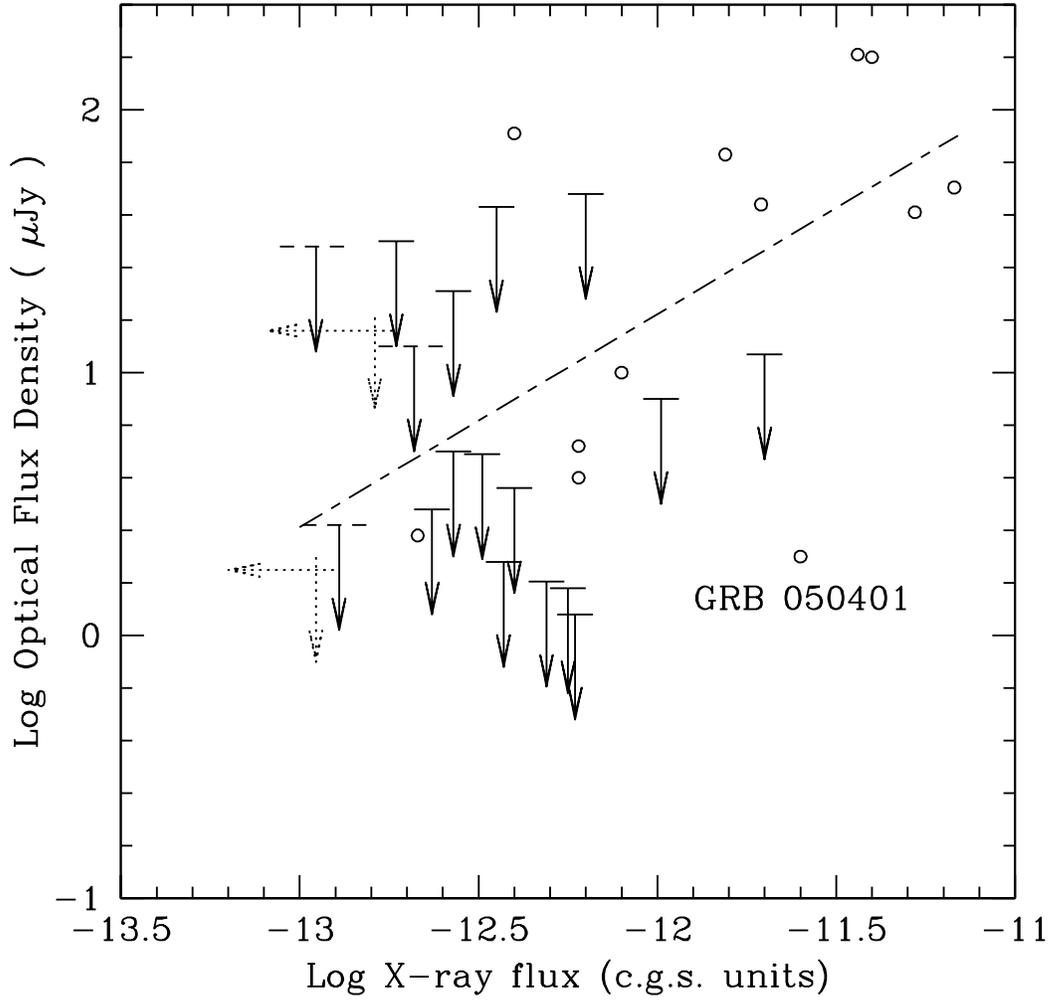}
 \caption{Optical vs. X-ray flux of GRB~050401 afterglow compared to
{\it BeppoSAX\/} GRBs. Open Circles: GRBs with optical counterpart.
Solid arrows: GRBs without optical counterpart. Dashed arrows:
doubtful afterglows. Dotted arrows: upper limits. Short-long dashed
line: best fit of optical vs. X-ray flux (adapted from De Pasquale
et al. 2003).}
\label{f5}
\end{figure*}

\end{document}